\documentclass[a4paper,12pt]{article}
\begin{document}
\begin{center}
\title{}
\textbf{Isospin Mass Differences of Heavy Baryons}\\
\bigskip
\bigskip
H. Fritzsch\\
\medskip
Physics Department\\
\smallskip
University of Munich, Germany
\end{center}
\bigskip
\begin{abstract}
\medskip
We discuss the mass differences for isospin multiplets of the charmed and b-flavored baryons.
The mass of the neutral b-flavored sigma baryon, which is not yet measured, is calculated. We point out,
that the measurements of the mass differences between the charmed sigma and chi baryons might be wrong.
\end{abstract}
\newpage
The mass splittings within isospin multiplets of the hadrons originate from the mass difference of the up and down quarks and from
electromagnetic effects. We study the heavy baryons, consisting of a c-quark or b-quark and two light quarks, in a simple quark model. The masses of the
heavy baryons, we should like to consider, are as follows (ref.1):
\begin{center}
$ \Sigma^{++}_{c}: \;\;\; 2454.02 \pm 0.18 \; MeV $\\
\medskip
$ \Sigma^{+}_{c}\;\:: \;\;\; 2452.90 \pm 0.40 \; MeV $\\
\medskip
$ \Sigma^{0}_{c}\;\:: \;\;\; 2453.76 \pm 0.18 \; MeV $
\begin{equation}
\Xi^{+}_{c}\;\:: \;\;\; 2467.90 \pm 0.40 \; MeV
\end{equation}
$ \Xi^{0}_{c}\;\:: \;\;\; 2471.00 \pm 0.40 \; MeV $\\
\medskip
$ \Sigma^{+}_{b}\;\:: \;\;\; 5807.80 \pm 2.70 \; MeV $\\
\medskip
$ \Sigma^{-}_{b}\;\:: \;\;\; 5815.20 \pm 2.00 \; MeV $
\end{center}

We note that the mass of the neutral baryon $ \Sigma_b $ has
not been measured. The following mass differences have been measured explicitly:
\begin{center}
$ m(\Sigma^{++}_{c})-m(\Sigma^{0}_{c})=0.27\pm 0.11 \; MeV $
\begin{equation}
m(\Sigma^{+}_{c})-m(\Sigma^{0}_{c})=-0.9\pm 0.40 \; MeV
\end{equation}
$ m(\Xi^{0}_{c})-m(\Xi^{+}_{c})=3.10\pm 0.50 \; MeV $
\end{center}

The isospin splittings of heavy baryons were studied previously invarious quark models
(ref. 2,3,4,5,6,7,8,9) and in chiral perturbation theory (ref. 10).
In the quark model the wave functions of the heavy baryons are given
by:
\begin{center}
$\Sigma^{++}_{c}:\;\:(cuu)\;\;\;\;\;\Sigma^{+}_{c}:\;\:(cud)\;\;\;\;\;\Sigma^{0}_{c}:\;\:(cdd)$
\begin{equation}
\Xi^{+}_{c}:\;\:(csu)\;\;\;\;\;\Xi^{0}_{c}:\;\:(csd)
\end{equation}
$\Sigma^{+}_{b}:\;\:(buu)\;\;\;\;\;\Sigma^{0}_{b}:\;\:(bud)\;\;\;\;\;\Sigma^{-}_{b}:\;\:(bdd)$
\end{center}

In the limit, in which the u- and d-quarks have no mass
and in which the electromagnetic interaction is turned off, we
describe the masses by parameters C, $ C^s $ and B:
\begin{center}
$ m(\Sigma^{++}_{c})=m(\Sigma^{+}_{c})=m(\Sigma^{0}_{c})=C $
\begin{equation}
m(\Xi^{+}_{c})=m(\Xi^{0}_{c})=C^s
\end{equation}
$ m(\Sigma^{+}_{b})=m(\Sigma^{0}_{b})=m(\Sigma^{-}_{b})=B $
\end{center}

In the next step we introduce the masses of the light quarks and
the electromagnetic interaction. We do not specify the details of the electromagnetic
interaction, which will depend on the internal structure of the baryons. We take into
account only the electric charges of the quarks. The mass shift caused by the electromagnetic
interaction is parametrized by a parameter A, which depends on the bound state structure of the
baryons and can not be calculated. We find:
\begin{center}
$m(\Sigma^{++}_{c})=C+2m_u+A\cdot(\frac{2}{3})^2\cdot3=C+2m_u+\frac{4}{3}A$\\
\bigskip
$m(\Sigma^{+}_{c})=C+m_u+m_d+A[(\frac{2}{3})(-\frac{1}{3})\cdot2+(\frac{2}{3})^2]=C+m_u+m_d$\\
\bigskip
$m(\Sigma^{0}_{c})=C+2m_d+A[(\frac{2}{3})(-\frac{1}{3})\cdot2+(-\frac{1}{3})^2]=C+2m_d-\frac{1}{3}A$
\begin{equation}
m(\Xi^{+}_{c})=C^s+m_u+A[(\frac{2}{3})(-\frac{1}{3})\cdot2+(\frac{2}{3})^2]=C^s+m_u
\end{equation}
$m(\Xi^{0}_{c})=C^s+m_d+A[(\frac{2}{3})(-\frac{1}{3})\cdot2+(-\frac{1}{3})^2]=C^s+m_d-\frac{1}{3}A$\\
\bigskip
$m(\Sigma^{+}_{b})=B+2m_u+A[(\frac{2}{3})(-\frac{1}{3})\cdot2+(\frac{2}{3})^2]=B+2m_u$\\
\bigskip
$m(\Sigma^{0}_{b})=B+m_u+m_d+A[(\frac{2}{3})(-\frac{1}{3})\cdot2+(-\frac{1}{3})^2]=B+m_u+m_d -\frac{1}{3}A$\\
\bigskip
$ m(\Sigma^{-}_{b})= B+2m_d+A(-\frac{1}{3})^2\cdot3=B+2m_d+\frac{1}{3}A$
\end{center}

Here $m_u$ denotes the matrix element $<baryon\; \vert m_u\overline{u}u\vert \; baryon>$, etc. We obtain the following relation:
\begin{equation}
m(\Sigma^{++}_{c})+m(\Sigma^{0}_{c})-2m(\Sigma^{+}_{c})=m(\Sigma^{+}_{b})+m(\Sigma^{-}_{b})-2m(\Sigma^{0}_{b})
\end{equation}

It is unknown, whether this relation is fulfilled by the experimental data, since the mass of the $ \Sigma^{0}_{b} $ is unknown.

We calculate the mass differences inside the isospin multiplets
in terms of the parameter A and the mass difference between the
d-quarks and the u-quarks, which we denote by x.
\begin{center}
$ m(\Sigma^{++}_{c}) - m (\Sigma^{0}_{c}) = - 2 x + \frac{5}{3} A $\\
\bigskip
$ m(\Sigma^{+}_{c}) - m (\Sigma^{0}_{c}) = - x + \frac{1}{3} A $
\begin{equation}
m(\Xi^{0}_{c}) - m (\Xi^{+}_{c}) = x - \frac{1}{3} A
\end{equation}
$ m(\Sigma^{+}_{b}) - m (\Sigma^{0}_{b}) = - x + \frac{1}{3} A $\\
\bigskip
$ m(\Sigma^{+}_{b}) - m (\Sigma^{-}_{b}) = - 2 x - \frac{1}{3} A $\\
\bigskip
$ m(\Sigma^{-}_{b}) - m (\Sigma^{0}_{b}) = x + \frac{2}{3} A $
\end{center}

In particular the mass of the neutral baryon $ \Sigma_b $ is determined
by the following relation:
\begin{equation}
m(\Sigma^{0}_{b})=m(\Sigma^{+}_{b})+m(\Xi^{0}_{c})-m(\Xi^{+}_{c})=5810.9\pm2.8\;\;MeV
\end{equation}
According to relation (7) the following mass differences should be equal:
\begin{equation}
m(\Sigma^{+}_{c}) - m (\Sigma^{0}_{c}) = m (\Xi^{+}_{c})-m(\Xi^{0}_{c})
\end{equation}

The experiments give for the first mass difference $ -0.9 \pm 0.4\; MeV$, for the second one $ -3.1 \pm 0.5 \; MeV $.
The results are not consistent. Either our model is wrong, or
the experiments are wrong. Further experiments might give values
for the two mass differences, which are consistent with relation(9).

We proceed to determine the values of A and $ m_d - m_u $.
The observed mass differences can be described rather well
by the following values:
\begin{center}
A = 3.2 MeV \\
\medskip
$ x = m_d - m_u = 2.5 \; MeV $.
\end{center}
We find, using these values:
\begin{center}
\begin{itemize}
\item [I.] $ m(\Sigma^{++}_{c}) - m (\Sigma^{0}_{c}) = 0.33 \;\;MeV $
\item[II.]$m(\Sigma^{+}_{c})-m(\Sigma^{0}_{c})=-1.43\;\;MeV\;\;=m(\Xi^{+}_{c})-m(\Xi^{0}_{c})$
\item [III.] $ m(\Sigma^{+}_{b}) - m (\Sigma^{-}_{b}) = -6.07 \;\;MeV $
\end{itemize}
\end{center}
The experimental values for these mass differences are:
\begin{itemize}
\item [I.]   $ 0.27 \pm 0.11 \; MeV $
\item [II.]  $ -0.9 \pm 0.4 MeV  and -3.1 \pm 0.5 \; MeV $
\item [III.] $ -7.4 \pm 3.3 \; MeV $
\end{itemize}

We conclude: We can describe the mass splittings inside the isospin multiplets of the heavy baryons in a simple quark model
rather well. But the discrepancies between the theoretical estimates
and the measurements should be resolved. Especially the relation(9) should be valid, implying that the two mass differences were not
measured correctly.
\newpage
References
\begin{enumerate}
\item  C. Amsler et al. (Particle Data group), Physl. Lett. B667(2008)1.
\item  K. P. Kiwari, C. P. Singh and M. P. Khanna, Phys.Rev.D31(1985)642
\item  S. Capstick, Phys. Rev. D 36 (1987) 2800
\item  L. H. Chan, Phys. Rev. D 31 (1985) 204
\item  W. Y. P. Hwang and D. B. Lichtenberg, Phys. Rev. D 35 (1987)3526
\item  R. C. Verma and S. Srivastava, Phys. Rev. D 38 (1988) 1623
\item  R. E. Cutkosky and P. Geiger, Phys. Rev. D 48 (1993) 1315
\item  K. Varga et al., Physl Rev. D 59 (1999) 014012
\item  B. Silvestre-Brac et al., J. Phys. G 29 (2003) 2685
\item  F. Guo, C. Hanhart and U. Meissner, arXiv:hep-ph/8092359
\end{enumerate}
\end{document}